\documentstyle[manuscript,eqsecnum,aps]{revtex}
\begin{document}

\tighten

\preprint{HEP-01-13}

\title{A Class of Exactly-Solvable Eigenvalue Problems}

\author{Carl M. Bender\cite{bye1} and Qinghai Wang\cite{bye2}}

\address{Department of Physics, Washington University, St. Louis, MO 63130, USA}

\date{\today}

\maketitle

\begin{abstract}
The class of differential-equation eigenvalue problems $-y''(x)+x^{2N+2}y(x)=x^N
Ey(x)$ ($N=-1,0,1,2,3,\ldots$) on the interval $-\infty<x<\infty$ can be solved
in closed form for all the eigenvalues $E$ and the corresponding eigenfunctions
$y(x)$. The eigenvalues are all integers and the eigenfunctions are all
confluent hypergeometric functions. The eigenfunctions can be rewritten as
products of polynomials and functions that decay exponentially as $x\to\pm
\infty$. For odd $N$ the polynomials that are obtained in this way are new and
interesting classes of orthogonal polynomials. For example, when $N=1$, the
eigenfunctions are orthogonal polynomials in $x^3$ multiplying Airy functions of
$x^2$. The properties of the polynomials for all $N$ are described in detail.
\end{abstract}


\section{Introduction}
\label{sec1}

In this paper we consider the class of differential-equation eigenvalue problems
\begin{equation}
-y''(x)+x^{2N+2}y(x)=x^NEy(x)\qquad(N=-1,0,1,2,3,\ldots)
\label{e1}
\end{equation}
on the interval $-\infty<x<\infty$. The eigenfunction $y(x)$ is required to obey
the boundary conditions that $y(x)$ vanish exponentially rapidly as $x\to\pm
\infty$. For each integer $N\geq-1$, it is possible to solve these eigenvalue
problems in closed form. The eigenvalues are all integers and the associated
eigenfunctions are all confluent hypergeometric functions. Furthermore, all the
eigenfunctions for each value of $N$ can be written as the product of a
polynomial and a given function that vanishes exponentially for large $|x|$. The
classes of polynomials that are obtained in this way are orthogonal and for odd
$N$ are apparently new and have interesting mathematical properties.

The eigenvalue problem (\ref{e1}) discussed in this paper arises in many
contexts. In classical physics a perturbative technique called boundary-layer
theory has been developed to find approximate solutions to boundary-value
problems of the form
\begin{equation}
\epsilon w''(x)+a(x)w'(x)+b(x)w(x)=0,\qquad w(-1)=A,~w(1)=B,
\label{ebl1}
\end{equation}
where $\epsilon$ is treated as a small parameter. Problems of this sort appear
in the study of fluid-flow problems in various geometries. Perturbative
treatments of this equation are usually quite straightforward \cite{BO}.
However, there is a particularly difficult special case of (\ref{ebl1}) that may
occur when there is a {\it resonant} internal boundary layer. Suppose that
$a(0)=0$, so that there is an internal boundary layer at $x=0$. Suppose further
that near $x=0$, $a(x)\sim\alpha x$ and $b(x)\sim\beta$. Then, near $x=0$ the
differential equation in (\ref{ebl1}) is approximated by
\begin{equation}
\epsilon w''(x)+\alpha xw'(x)+\beta w(x)=0.
\label{ebl2}
\end{equation}
The Gaussian change of variables $w(x)=e^{-ax^2/(4\epsilon)}y(x)$ converts
(\ref{ebl2}) to Schr\"odinger form:
\begin{equation}
-y''(x)+{\alpha^2\over4\epsilon^2}x^2y(x)={2\beta-\alpha\over2\epsilon}y(x).
\label{ebl3}
\end{equation}
Apart from a scaling, this equation is the $N=0$ case of (\ref{e1}). It
describes the quantum harmonic oscillator, and eigenvalues occur when the
parameters $\alpha$ and $\beta$ satisfy
\begin{equation}
\beta=(n+1)\alpha\qquad(n=0,1,2,3,\ldots).
\label{ebl4}
\end{equation}
When the parameters $\alpha$ and $\beta$ are related in this fashion, the
internal boundary layer is said to be {\it resonant}. Unlike conventional
boundary layers, a resonant boundary layer is not narrow; that is, its thickness
is not small when $\epsilon$ is small. As a result it is particularly difficult
to treat the resonant case using ordinary boundary-layer methods \cite{BO}.

More generally, if $a(x)\sim\alpha x^{N+1}$ and $b(x)\sim\beta x^N$ when $x$
is near $0$, we obtain the differential equation in (\ref{e1b}). This equation
is then converted to the differential equation in (\ref{e1}) by the exponential
change of variables in (\ref{e1a}). Thus, the eigenvalue problem in (\ref{e1})
characterizes the {\it general} resonant case in the theory of internal boundary
layers. The special case of the harmonic oscillator discussed above occurs
when $N=0$.

The eigenvalue problem also arises in the context of supersymmetric quantum
mechanics and quasi-exactly solvable models. It appears, for example, in the
recent work of Voros \cite{VOROS} and Dorey {\it et al} \cite{DDT}. The same
differential equation was also considered by Znojil \cite{Z} but with boundary
conditions imposed on a semi-infinite interval. The quantum problem in
(\ref{e1}) may be thought of as a peculiar inverse approach to quasi-exact
solvability. Ordinarily, in this field one tries to construct potentials for
which a finite number of eigenvalues of the spectrum can be calculated exactly
and in closed form, while the remaining part of the spectrum remains
analytically intractable. The problem in (\ref{e1}) is to construct potentials
$V(x)$ for which there is an eigenvalue that is exactly zero. The zero
eigenvalue may or may not be the ground-state energy of the potential $V(x)=
x^{2N+2}-x^NE$ that has been constructed. In this paper we will see that
the case of odd-integer $N$ is much more interesting than the even-$N$ case.

The eigenvalue problem in (\ref{e1}) is especially interesting because, as we
show in Sec.~\ref{sec2}, leading-order WKB theory (physical optics) gives the
exact spectrum $E$ for all odd $N$ and almost the exact answer for even $N$.

This paper is organized as follows. In Sec.~\ref{sec2} we give the exact
solution to the eigenvalue problem in (\ref{e1}). We show that the eigenvalues
$E$ are all integers for each value of $N=-1,0,1,2,3,\ldots$ and that the
corresponding eigenfunctions are all confluent hypergeometric functions. In
Sec.~\ref{sec3} we examine the eigenfunctions for even-integer $N$. For this
case the eigenspectrum is positive and the $n$th eigenfunction has definite
parity. The $n$th eigenfunction has the form of a polynomial of degree $n$
and of argument $x^{N+2}$ multiplied by the exponential $\exp(-x^{N+2})$, which
decays as $x\to\pm\infty$. The polynomials are generalized Laguerre polynomials.
The polynomials for even $n$ form an orthogonal set and the polynomials for
odd $n$ form a different orthogonal set. In Sec.~\ref{sec4}, we study the
eigenfunctions for odd-integer $N$. For this case the spectrum of eigenvalues
$E$ ranges from $-\infty$ to $\infty$ and the $n$th eigenfunction does not
exhibit definite parity. For each $N$ the $n$th eigenfunction has the general
form $xA_N(x^{N+1})P_n(x^{N+2})+A_N'(x^{N+1})Q_n(x^{N+2})$. Here, $P_n(z)$ and
$Q_n(z)$ are polynomials of degree $n$ that satisfy the same recursion relation
but have different initial conditions. The functions $A_N(z)$ are independent of
$n$ and are generalized Airy functions that obey the differential equation
$A_N''(z)=z^{2/(N+1)}A_N(z)$. When $N=1$, the function $A_1(z)$ is the
conventional Airy function ${\rm Ai}(z)$. In Sec.~\ref{sec5} we consider the
special cases $N=-1$ and $N=1$ (the Airy case).

We emphasize that the eigenvalues that are obtained in this paper are not the
energies of a conventional Schr\"odinger equation. However, the eigenfunctions
that are obtained might well be useful for solving some conventional
Schr\"odinger equations. For example, it might be useful to express the
eigenfunctions of the pure anharmonic oscillator problem, $-y''(x)+x^4y(x)=
Ey(x)$ as linear combinations of the eigenfunctions found in this paper. The
work done in this paper suggests that it may well be advantageous to expand the
function $y(x)$ as $P(x){\rm Ai}(x^2)+Q(x){\rm Ai}'(x^2)$, where $P$ and $Q$ are
series in powers of $x$.

\section{Exact Solution of the Eigenvalue Problem}
\label{sec2}

We solve for the eigenvalues of the differential equation (\ref{e1}) by
converting it to a confluent hypergeometric equation and then imposing the
boundary conditions. We begin by making the substitution
\begin{equation}
y(x)=e^{-{1\over N+2}x^{N+2}}w(x).
\label{e1a}
\end{equation}
The function $w(x)$ then satisfies the differential equation
\begin{equation}
w''(x)-2x^{N+1}w'(x)+\beta x^N w(x)=0,
\label{e1b}
\end{equation}
where
\begin{equation}
\beta\equiv E-N-1.
\label{e1c}
\end{equation}

There are now two cases to consider, $N$ even and $N$ odd. Suppose first that
$N$ is even. There are two linearly independent solutions to the differential
equation
(\ref{e1b}):
\begin{equation}
w(x)={}_1F_1\left(-{\beta\over2(N+2)},1-{1\over N+2};{2\over N+2}x^{N+2}\right)
\label{e1d}
\end{equation}
and
\begin{equation}
w(x)=x\,{}_1F_1\left({1\over N+2}-{\beta\over2(N+2)},1+{1\over N+2};{2
\over N+2}x^{N+2}\right).
\label{e1e}
\end{equation}

When the first parameter of the confluent hypergeometric function
is a negative integer, its Taylor series truncates to a Laguerre polynomial:
\begin{equation}
{}_1F_1(-n,c;t)=n!\,{\Gamma(c)\over\Gamma(c-n)}\,{\rm L}_n^{(c-1)}(t).
\label{e1f}
\end{equation}
Thus, for the solution in (\ref{e1d}) we obtain
\begin{equation}
{\beta_n\over2(N+2)}=n\qquad(n=0,1,2,3,\ldots),
\label{e1g}
\end{equation}
or
\begin{equation}
E_n=2n(N+2)+N+1\qquad(n=0,1,2,3,\ldots).
\label{e1h}
\end{equation}
The corresponding eigenfunctions have even parity. For the solution in
(\ref{e1e}) we obtain
\begin{equation}
-{1\over N+2}+{\beta_n\over2(N+2)}=n\qquad(n=0,1,2,3,\ldots),
\label{e1i}
\end{equation}
or
\begin{equation}
E_n=2n(N+2)+N+3\qquad(n=0,1,2,3,\ldots).
\label{e1j}
\end{equation}
The corresponding eigenfunctions have odd parity.

When $N$ is odd, the solution to the differential equation (\ref{e1b}) that is
well behaved as $x\to\infty$ is a particular linear combination of the solutions
in (\ref{e1d}) and (\ref{e1e}) known as a confluent hypergeometric function of
the second kind:
\begin{eqnarray}
w(x)&=&{\Gamma\left({1\over N+2}\right)\over\Gamma\left({1\over N+2}-
{\beta\over2(N+2)}\right)}\,
{}_1F_1\left(-{\beta\over2(N+2)},1-{1\over N+2};{2\over N+2}x^{N+2}\right)
\nonumber\\
&+&{\Gamma\left(-{1\over N+2}\right)\over\Gamma\left(-{\beta\over2(N+2)}\right)}
\,x\,{}_1F_1\left({1\over N+2}-{\beta\over2(N+2)},1+{1\over N+2};{2
\over N+2}x^{N+2}\right).
\label{e1k}
\end{eqnarray}
Note that as $x\to+\infty$, the function $y(x)$ in (\ref{e1a}) vanishes
exponentially. However, as $x\to-\infty$, the exponential factor in (\ref{e1a})
grows and we have
\begin{eqnarray}
y(x)&\sim& e^{-{1\over N+2}x^{N+2}}\left(-{2\over
N+2}x^{N+2}\right)^{\beta\over2(N+2)}\nonumber\\
&&\times\left[
{\Gamma\left({1\over N+2}\right)\Gamma\left(1-{1\over N+2}\right)\over
\Gamma\left({1\over N+2}-{\beta\over2(N+2)}\right)\Gamma\left(1-{1\over N+2}
+{\beta\over2(N+2)}\right)}
-{\Gamma\left(-{1\over N+2}\right)\Gamma\left(1+{1\over N+2}\right)\over
\Gamma\left(-{\beta\over2(N+2)}\right)\Gamma\left(1+{\beta\over2(N+2)}\right)}
\right]\nonumber\\
&&\times \left[1+{\rm O}\left({1\over |x|^{N+2}}\right)\right].
\label{e1l}
\end{eqnarray}
Because $\exp\left(-{1\over N+2}x^{N+2}\right)$ grows exponentially as
$x\to-\infty$, the only way to satisfy the boundary condition as $x\to-\infty$
is for the expression in square brackets to vanish. The expression in square
brackets simplifies to
\begin{equation}
{ \sin\left[\left({1\over N+2}-{\beta\over2(N+2)}\right)\pi\right]
-\sin{\beta \pi\over2(N+2)} \over \sin\left({\pi\over N+2}\right)}.
\label{e1m}
\end{equation}
Hence,
\begin{equation}
{\beta_n\over N+2}={1\over N+2}+2n\qquad(n=0,\pm1,\pm2,\pm3,\ldots),
\label{e1n}
\end{equation}
and thus we obtain the eigenvalues
\begin{equation}
E_n=(2n+1)(N+2)\qquad(n=0,\pm1,\pm2,\pm3,\ldots).
\label{e1o}
\end{equation}

It is interesting that a leading-order WKB analysis (the physical optics
approximation) of (\ref{e1}) gives the exact eigenvalues for odd $N$ and almost
the exact eigenvalues when $N$ is even. Consider the following two-turning-point
time-independent Schr\"odinger equation boundary-value problem
\begin{equation}
-y''(x)+Q(x)y(x)=0,\qquad y(\pm\infty)=0.
\label{ew1}
\end{equation}
Ordinarily, $Q(x)=V(x)-E$, where $V(x)$ is the potential and $E$ is the energy.
In the physical-optics approximation, the condition for a solution to this
problem to exist is
\begin{equation}
\int_A^B dx\,\sqrt{-Q(x)}=\left(n+{1\over2}\right)\pi\qquad(n=0,1,2,3,\ldots),
\label{ew2}
\end{equation}
where the turning points $A$ and $B$ satisfy $Q(A)=Q(B)=0$. If we apply the
quantization condition (\ref{ew2}) to (\ref{e1}), where $Q(x)=x^{2N+2}-Ex^N$, we
obtain
\begin{equation}
\int_0^B dx\,\sqrt{Ex^N-x^{2N+2}}=\left(n+{1\over2}\right)\pi\qquad
(n=0,1,2,3,\ldots),
\label{ew3}
\end{equation}
where we assume without loss of generality that $E$ is positive. The turning
point $B$ satisfies $B^{N+2}=E$. This integral can be evaluated exactly as a
beta function and we obtain
\begin{equation}
E=(2n+1)(N+2).
\label{ew4}
\end{equation}
This is precisely the result in (\ref{e1o}) for $N$ odd. Also, it is nearly
the results for even $N$ in (\ref{e1h}) and (\ref{e1j}), which can be combined
to read 
\begin{equation}
E=(2n+1)(N+2)\pm 1.
\label{ew5}
\end{equation}

A striking property of the eigenvalue problem (\ref{e1}) is that for even $N$
the eigenvalues are positive but for odd $N$ the eigenvalues are both positive
and negative. This is reminiscent of the difference between the Klein-Gordon
equation for bosons, which has positive-energy states only, and the Dirac
equation for fermions, which has positive-energy states (electrons) and
negative-energy states (positrons or holes).

In the next two sections we describe the two cases $N$ odd and $N$ even in
greater depth. We consider the simpler case of even $N$ in Sec.~\ref{sec3} and
turn to the more interesting case of odd $N$ in Secs.~\ref{sec4} and \ref{sec5}.

\section{Eigenvalue Problem for Even $N$}
\label{sec3}

When $N$ is even, all the eigenvalues are positive and the eigenfunctions have
either even or odd parity. The even-parity eigenfunctions have the form
\begin{equation}
y_{2n}(x)=e^{-{1\over N+2}x^{N+2}}{\rm L}_n^{\left(-{1\over N+2}\right)}
\left({2\over N+2}x^{N+2}\right)\qquad(n=0,1,2,3,\ldots)
\label{e2}
\end{equation}
and the corresponding eigenvalues are
\begin{equation}
E_{2n}=2n(N+2)+N+1,
\label{e3}
\end{equation}
where ${\rm L}_n^{(\alpha)}$ is the generalized Laguerre polynomial. The
odd-parity eigenfunctions are
\begin{equation}
y_{2n+1}(x)=e^{-{1\over N+2}x^{N+2}}x\,{\rm L}_n^{\left({1\over N+2}\right)}
\left({2\over N+2}x^{N+2}\right)\qquad(n=0,1,2,3,\ldots)
\label{e4}
\end{equation}
and the corresponding eigenvalues are
\begin{equation}
E_{2n+1}=2n(N+2)+N+3.
\label{e5}
\end{equation}

Note that the eigenfunctions have the form of a decaying exponential multiplying
a polynomial. For the even-parity solutions we can write the polynomial as a
monic\footnote{The term {\it monic} means that the coefficient of the highest
power in the polynomial is 1.} polynomial $p_n(z)$ in the variable $z=4x^{N+2}$:
\begin{equation}
p_n(z) = (-1)^n n!\,[2(N+2)]^n\,{\rm L}_n^{\left(-{1\over N+2}\right)}\left(
{z\over2(N+2)}\right)
\label{e6}
\end{equation}
and for the odd-parity solutions we have
\begin{equation}
q_n(z) = (-1)^n n!\,[2(N+2)]^n\,{\rm L}_n^{\left({1\over N+2}\right)}\left(
{z\over2(N+2)}\right).
\label{e7}
\end{equation}

These polynomials satisfy very similar recurrence relations
\begin{eqnarray}
p_{n+1}(z)&=&[z-2(N+2)(2n+1)+2]p_n(z)-2(N+2)n[2(N+2)n-2]p_{n-1}(z),\nonumber\\
q_{n+1}(z)&=&[z-2(N+2)(2n+1)-2]q_n(z)-2(N+2)n[2(N+2)n+2]q_{n-1}(z),
\label{e8}
\end{eqnarray}
where the initial conditions are
\begin{eqnarray}
p_0(z)=1, && \qquad p_1(z)=z-2N-2,\nonumber\\
q_0(z)=1, && \qquad q_1(z)=z-2N-6.
\label{e9}
\end{eqnarray}
The polynomials $p_n(z)$ and $q_n(z)$ also obey similar differential equations
\begin{eqnarray}
2(N+2)zp_n''(z)+(2N+2-z)p_n'(z)+np_n(z)&=&0,\nonumber\\
2(N+2)zq_n''(z)+(2N+6-z)q_n'(z)+nq_n(z)&=&0,
\label{e10}
\end{eqnarray}
and differential relations
\begin{eqnarray}
2(N+2)zp_n'(z)+p_{n+1}(z)+[2(N+2)(n+1)-2-z]p_n(z)&=&0,\nonumber\\
2(N+2)zq_n'(z)+q_{n+1}(z)+[2(N+2)(n+1)+2-z]q_n(z)&=&0.
\label{e11}
\end{eqnarray}
The generating functions for these polynomials are also quite similar:
\begin{eqnarray}
G_p(z,t)&\equiv&\sum_{n=0}^\infty {(-1)^nt^n\over n!}p_n(z)=[1-2(N+2)t]^{-1+{1
\over N+2}}\,e^{zt\over2(N+2)t -1},\nonumber\\
G_q(z,t)&\equiv&\sum_{n=0}^\infty {(-1)^nt^n\over n!}q_n(z)=[1-2(N+2)t]^{-1-{1
\over N+2}}\,e^{zt\over2(N+2)t -1}.
\label{e12}
\end{eqnarray}

The polynomials $p_n(z)$ and $q_n(z)$ are separately orthogonal:
\begin{eqnarray}
\int_0^\infty dz\,w_p(z)p_m(z)p_n(z)&=&[2(N+2)]^{2n}n!\,{\Gamma\left(n+1-{1\over
N+2}\right)\over\Gamma\left(1-{1\over N+2}\right)}\,\delta_{m,n},\nonumber\\
\int_0^\infty dz\,w_q(z)q_m(z)q_n(z)&=&[2(N+2)]^{2n}n!\,{\Gamma\left(n+1+{1\over
N+2}\right)\over\Gamma\left(1+{1\over N+2}\right)}\,\delta_{m,n},
\label{e13}
\end{eqnarray}
where the weight functions $w_p(z)$ and $w_q(z)$ are given by
\begin{eqnarray}
w_p(z)&=&e^{-{z\over2(N+2)}}z^{-{1\over N+2}}[2(N+2)]^{{1\over N+2}-1}
{1\over\Gamma\left(1-{1\over N+2}\right)},\nonumber\\
w_q(z)&=&e^{-{z\over2(N+2)}}z^{1\over N+2}[2(N+2)]^{-{1\over N+2}-1}
{1\over\Gamma\left(1+{1\over N+2}\right)}.
\label{e14}
\end{eqnarray}
The moments of these weight functions are
\begin{eqnarray}
a_n^{(p)}&\equiv&\int_0^\infty dz\,w_p(z)z^n=[2(N+2)]^n{\Gamma\left(n+1-{1\over
N+2}\right)\over\Gamma\left(1-{1\over N+2}\right)},\nonumber\\
a_n^{(q)}&\equiv&\int_0^\infty dz\,w_q(z)z^n=[2(N+2)]^n{\Gamma\left(n+1+{1\over
N+2}\right)\over\Gamma\left(1+{1\over N+2}\right)}.
\label{e15}
\end{eqnarray}
The (divergent) power series constructed from these moments have particularly
simple continued fraction expansions in which the continued-fraction
coefficients are all integers:
\begin{eqnarray}
\sum_{n=0}^\infty a_n^{(p)}t^n&=&
\displaystyle{1\over 1-
\displaystyle{[2(N+2)-2]t\over 1-
\displaystyle{[2(N+2)]t\over 1-
\displaystyle{[4(N+2)-2]t\over 1-
\displaystyle{[4(N+2)]t\over 1-
\displaystyle{[6(N+2)-2]t\over 1-
\displaystyle{[6(N+2)]t\over 1-\cdots}}}}}}},\nonumber\\
\sum_{n=0}^\infty a_n^{(q)}t^n&=&
\displaystyle{1\over 1-
\displaystyle{[2(N+2)+2]t\over 1-
\displaystyle{[2(N+2)]t\over 1-
\displaystyle{[4(N+2)+2]t\over 1-
\displaystyle{[4(N+2)]t\over 1-
\displaystyle{[6(N+2)+2]t\over 1-
\displaystyle{[6(N+2)]t\over 1-\cdots}}}}}}}.
\label{e16}
\end{eqnarray}

We illustrate these general results for the two special cases $N=0$ and $N=2$.

\noindent{\bf Special Case $N=0$: The Harmonic Oscillator.} For this case the
eigenvalues in (\ref{e3}) are $E=1,~3,~5,~7,~\ldots$ and the polynomials
$p_n(z)$ and $xq_n(z)$ in (\ref{e6}) and (\ref{e7}) coalesce to become the
standard Hermite polynomials ${\rm H}_n(x)$:
\begin{eqnarray}
p_n(z)&=&{\rm H}_{2n}(x)=(-1)^n 2^{2n} n!\,{\rm L}_n^{(-1/2)}(x^2),\nonumber\\
xq_n(z)&=&{\rm H}_{2n+1}(x)=(-1)^n 2^{2n+1} n!\,x{\rm L}_n^{(1/2)}(x^2).
\label{e17}
\end{eqnarray}

\noindent{\bf Special Case $N=2$.} For this case the eigenvalues in (\ref{e3})
are $E=3,~5,~11,~13,~19,~21,~\ldots$ and the first few monic polynomials
$p_n(x)$ and $q_n(x)$ in (\ref{e6}) and (\ref{e7}) are
\begin{eqnarray}
p_0(z)&=&1,\nonumber\\
p_1(z)&=&z-6,\nonumber\\
p_2(z)&=&z^2-28z+84,\nonumber\\
p_3(z)&=&z^3-66z^2+924z-1848,\nonumber\\
p_4(z)&=&z^4-120z^3+3960z^2-36960z+55440,\nonumber\\
q_0(z)&=&1,\nonumber\\
q_1(z)&=&z-10,\nonumber\\
q_2(z)&=&z^2-36z+180,\nonumber\\
q_3(z)&=&z^3-78z^2+1404z-4680,\nonumber\\
q_4(z)&=&z^4-136z^3+5304z^2-63648z+159120.
\label{e18}
\end{eqnarray}

\section{Eigenvalue Problem for Odd $N$}
\label{sec4}

For this case the eigenvalues are
\begin{equation}
E_n=(2n+1)(N+2)\qquad(n=0,\pm1,\pm2,\pm3,\ldots)
\label{e19}
\end{equation}
and the corresponding eigenfunctions $y_n(x)$, which do not have definite
parity, are confluent hypergeometric functions of the second kind:
\begin{equation}
y_n(x)=e^{-{1\over N+2}x^{N+2}}\,{\rm U}\left(-n-{1\over2(N+2)},1-{1\over
N+2};{2\over N+2}x^{N+2}\right).
\label{e20}
\end{equation}
Note that the boundary condition as $x\to+\infty$ is already satisfied and the
quantization comes from requiring that $y(x)\to0$ as $x\to-\infty$.

For each $N$ the eigenfunctions $y_n(x)$ in (\ref{e20}) can be expressed in
terms of what we will call {\it generalized Airy functions} $A_N(x)$ combined
with polynomials $P_n$ and $Q_n$ as follows:\footnote{To avoid confusion, for
odd $N$ we use upper-case notation $P_n$ and $Q_n$ to represent the polynomials;
we use lower-case notation $p_n$ and $q_n$ to represent the polynomials
associated with even $N$.}
\begin{eqnarray}
y_{-n-1}(x)&=&2^{-{N+1\over2(N+2)}}\,x
A_N\left[\left(2^{-{N+1\over2(N+2)}}\,x\right)^{N+1}\right] P_n(4x^{N+2})
\nonumber\\
&&\quad +A_N'\left[\left(2^{-{N+1\over2(N+2)}}\,x\right)^{N+1}\right]
Q_n(4x^{N+2})\qquad(n\geq0)
\label{e21a}
\end{eqnarray}
and
\begin{equation}
y_n(x)=y_{-n-1}(-x)\qquad(n\geq0).
\label{e21b}
\end{equation}

We define the {\it generalized Airy function} $A_N(x)$ as the solution to the
differential equation
\begin{equation}
A_N''(x)=x^{2\over N+1}A_N(x)
\label{e22}
\end{equation}
that decays exponentially as $x\to+\infty$. Note that when $N=1$, the function
$A_1(x)$ is just the conventional Airy function ${\rm Ai}(x)$. We can express
$A_N(x)$ in terms of the associated Bessel function $K_\nu(z)$ as follows:
\begin{equation}
A_N(x)={1\over2\pi}[4(N+1)]^{N+1\over2(N+2)}\sqrt{x\over N+2}
K_{N+1\over2(N+2)}\left({N+1\over N+2}x^{N+2\over N+1}\right).
\label{e23}
\end{equation}
With this choice the function $A_N(x)$ is normalized so that
\begin{equation}
\int_0^\infty dx\,A_N(x)={1\over\pi}\Gamma\left({N+1\over N+2}\right)\Gamma
\left({N+1\over2(N+2)}\right)2^{-{N+7\over2(N+2)}}(N+1)^{1\over N+2}(N+2)^{-{3
\over2(N+2)}}.
\label{e24}
\end{equation}
Note that this reduces to the standard result $\int_0^\infty dx\,{\rm Ai}(x)=
{1\over3}$ when $N=1$.

The polynomials $P_n(z)$ and $Q_n(z)$, where $z=4x^{N+2}$, both satisfy the
{\it same} recursion relation
\begin{eqnarray}
P_{n+1}(z)&=&[z+2(N+2)(2n+1)]P_n(z)-[2(N+2)n-1][2(N+2)n+1]P_{n-1}(z),\nonumber\\
Q_{n+1}(z)&=&[z+2(N+2)(2n+1)]Q_n(z)-[2(N+2)n-1][2(N+2)n+1]Q_{n-1}(z),
\label{e25}
\end{eqnarray}
but have different initial conditions
\begin{eqnarray}
P_0(z)=1, && \qquad P_1(z)=z+2N+5,\nonumber\\
Q_0(z)=1, && \qquad Q_1(z)=z+2N+3.
\label{e26}
\end{eqnarray}

The polynomials $P_n(z)$ and $Q_n(z)$ satisfy {\it coupled} second-order
differential equations
\begin{eqnarray}
4(N+2)zP_n''(z)+4(N+3)P_n'(z)+2zQ_n'(z)+Q_n(z)&=&(2n+1)P_n(z),\nonumber\\
4(N+2)zQ_n''(z)+4(N+1)Q_n'(z)+2zP_n'(z)+P_n(z)&=&(2n+1)Q_n(z),
\label{e27}
\end{eqnarray}
and {\it coupled} differential relations
\begin{eqnarray}
4(N+2)zP_n'(z)&=&2P_{n+1}(z)-zQ_n(z)-[4(N+2)(n+1)+2+z]P_n(z),\nonumber\\
4(N+2)zQ_n'(z)&=&2Q_{n+1}(z)-zP_n(z)-[4(N+2)(n+1)-2+z]Q_n(z).
\label{e28}
\end{eqnarray}
The generating functions for these polynomials are
\begin{eqnarray}
G_P(z,t)&\equiv&\sum_{n=0}^\infty {t^n\over n!}P_n(z)\nonumber\\
&=& [1-2(N+2)t]^{-1-{1 \over2(N+2)}}\,
{{}_1F_1\left(1+{1\over2(N+2)},1+{1\over N+2};{z\over2(N+2)[1-2(N+2)t]}\right)
\over {}_1F_1\left(1+{1\over2(N+2)},1+{1\over N+2};{z\over2(N+2)}\right)},
\label{e29a}
\end{eqnarray}
and
\begin{eqnarray}
G_Q(z,t)&\equiv&\sum_{n=0}^\infty {t^n\over n!}Q_n(z)\nonumber\\
&=&[1-2(N+2)t]^{-1+{1 \over2(N+2)}}\,
{{}_1F_1\left(1-{1\over2(N+2)},1-{1\over N+2};{z\over2(N+2)[1-2(N+2)t]}\right)
\over {}_1F_1\left(1-{1\over2(N+2)},1-{1\over N+2};{z\over2(N+2)}\right)}.
\label{e29b}
\end{eqnarray}

The polynomials $P_n(z)$ and $Q_n(z)$ obey identical-looking orthogonality
and normalization conditions
\begin{eqnarray}
\int_{-\infty}^\infty dx\,W_P(x)P_m(x)P_n(x)&=&
{1\over\pi}\sin{\pi\over2(N+2)}[2(N+2)]^{2n+1}
\Gamma\left(n+1-{1\over2(N+2)}\right)\nonumber\\
&&\quad\times \Gamma\left(n+1+{1\over2(N+2)}\right)
\,\delta_{m,n},\nonumber\\
\int_{-\infty}^\infty dx\,W_Q(x)Q_m(x)Q_n(x)&=&
{1\over\pi}\sin{\pi\over2(N+2)}[2(N+2)]^{2n+1}
\Gamma\left(n+1-{1\over2(N+2)}\right)\nonumber\\
&&\quad\times\Gamma\left(n+1+{1\over2(N+2)}\right)\,\delta_{m,n}.
\label{e30}
\end{eqnarray}
The weight functions $W_P(x)$ and $W_Q(x)$ are real and positive and are
expressible as principal-part integrals:
\begin{eqnarray}
W_P(x)&=&\int_{-\infty}^x ds\,{\cal P}\int_{-\infty}^\infty {dt\over t-s}
\ln\left[\sqrt{-t\over2(N+2)\pi}\,e^{-{t\over4(N+2)}}\,K_{N+3\over2(N+2)}
\left({-t\over4(N+2)}\right)\right],\nonumber\\
W_Q(x)&=&\int_{-\infty}^x ds\,{\cal P}\int_{-\infty}^\infty {dt\over t-s}
\ln\left[\sqrt{-t\over2(N+2)\pi}\,e^{-{t\over4(N+2)}}\,K_{N+1\over2(N+2)}
\left({-t\over4(N+2)}\right)\right],
\label{e31}
\end{eqnarray}
or, in terms of the generalized Airy functions $A_N(x)$,
\begin{eqnarray}
W_P(x)&=&\int_{-\infty}^x ds\,{\cal P}\int_{-\infty}^\infty {dt\over t-s}
\ln\left\{-\sqrt{2\pi}(-t)^{1\over2(N+2)}e^{-{t\over4(N+2)}}A_N'
\left[\left({-t\over4(N+1)}\right)^{N+1\over N+2}\right]\right\},\nonumber\\
W_Q(x)&=&\int_{-\infty}^x ds\,{\cal P}\int_{-\infty}^\infty {dt\over t-s}
\ln\left\{\sqrt{2\pi}(-t)^{1\over2(N+2)}e^{-{t\over4(N+2)}}A_N
\left[\left({-t\over4(N+1)}\right)^{N+1\over N+2}\right]\right\},
\label{e32}
\end{eqnarray}
where ${\cal P}$ indicates principal-part integration and the integral is
performed on the sheet for which $-1\equiv e^{-i\pi}$.

The moments of the weight functions $W_P(x)$ and $W_Q(x)$ are given by
\begin{equation}
a_n^{(P)}\equiv\int_{-\infty}^\infty dx\,x^n W_P(x)\quad{\rm and}\quad
a_n^{(Q)}\equiv\int_{-\infty}^\infty dx\,x^n W_Q(x).
\label{e33}
\end{equation}
The divergent power series constructed from these moments have remarkably simple
continued-fraction expansions in which the continued-fraction coefficients are
all integers:
\begin{eqnarray}
\sum_{n=0}^\infty a_n^{(P)}t^n&=&
\displaystyle{1\over 1-
\displaystyle{[2(N+2)+1]t\over 1-
\displaystyle{[2(N+2)-1]t\over 1-
\displaystyle{[4(N+2)+1]t\over 1-
\displaystyle{[4(N+2)-1]t\over 1-
\displaystyle{[6(N+2)+1]t\over 1-
\displaystyle{[6(N+2)-1]t\over 1-\cdots}}}}}}},\nonumber\\
\sum_{n=0}^\infty a_n^{(Q)}t^n&=&
\displaystyle{1\over 1-
\displaystyle{[2(N+2)-1]t\over 1-
\displaystyle{[2(N+2)+1]t\over 1-
\displaystyle{[4(N+2)-1]t\over 1-
\displaystyle{[4(N+2)+1]t\over 1-
\displaystyle{[6(N+2)-1]t\over 1-
\displaystyle{[6(N+2)+1]t\over 1-\cdots}}}}}}}.
\label{e34}
\end{eqnarray}

\section{Two Special Cases of the Odd-$N$ Eigenvalue Problem}
\label{sec5}

In this section we consider two interesting special cases of the odd-$N$
eigenvalue problem; namely, $N=-1$ and $N=1$.

\noindent{\bf Special case $N=-1$.} For this case Eq.~(\ref{e22}) is of course
not valid. However, the formula for the eigenvalues $E$ in (\ref{e19}) is still
valid and $E=\pm1,~\pm3,~\pm5,~\pm7,~\ldots$. The eigenfunctions $y_n(x)$ in
(\ref{e20}) are now {\it Bateman functions} $k_{E_n}(x)$:
\begin{equation}
y_n(x)=k_{E_n}(x)\equiv{2\over\pi}\int_0^{\pi/2}d\theta\,\cos(x\,\tan\theta-E_n
\theta).
\label{e35}
\end{equation}
Apart from an overall multiplicative constant, the solution can be written
in terms of associated Bessel functions combined with polynomials. For
negative eigenvalues
\begin{equation}
y_{-n-1}(x)=xK_0(x)P_n(4x)+xK_0'(x)Q_n(4x)\qquad(n\geq0),
\label{e36}
\end{equation}
and for positive eigenvalues
\begin{equation}
y_n(x)\equiv y_{-n-1}(-x)\qquad(n\geq0).
\label{e37}
\end{equation}
Note that the eigenfunctions are finite at the origin but that there is a branch
cut. For definiteness, we take the branch cut to run up the positive
imaginary-$x$ axis.

In terms of the variable $z=4x$ the first few polynomials $P_n(z)$ and $Q_n(z)$
are
\begin{eqnarray}
P_0(z)&=&1,\nonumber\\
P_1(z)&=&z+3,\nonumber\\
P_2(z)&=&z^2+9z+15,\nonumber\\
P_3(z)&=&z^3+19z^2+90z+105,\nonumber\\
P_4(z)&=&z^4+33z^3+321z^2+1050z+945,\nonumber\\
P_5(z)&=&z^5+51z^4+852z^3+5631z^2+14175z+10395,\nonumber\\
Q_0(z)&=&1,\nonumber\\
Q_1(z)&=&z+1,\nonumber\\
Q_2(z)&=&z^2+7z+3,\nonumber\\
Q_3(z)&=&z^3+17z^2+58z+15,\nonumber\\
Q_4(z)&=&z^4+31z^3+261z^2+582z+105,\nonumber\\
Q_5(z)&=&z^5+49z^4+756z^3+4209z^2+6927z+945.
\label{e38}
\end{eqnarray}

The polynomials $P_n(x)$ and $Q_n(x)$ satisfy the recursion relations in
(\ref{e25}) with $N=-1$, the coupled second-order differential equations
in (\ref{e27}) with $N=-1$, and the coupled differential relations in
(\ref{e28}) with $N=-1$. The generating functions for $P_n(z)$ and $Q_n(z)$ are
expressed in terms of Bateman functions $k_\nu$:
\begin{eqnarray}
G_P(z,t)&\equiv&\sum_{n=0}^\infty {t^n\over n!}P_n(z)=(1-2t)^{-3/2}\,
{e^{z\over4(1-2t)}\,k_{-3}\left({z\over4(1-2t)}\right)\over
e^{z/4}\,k_{-3}(z/4)}\nonumber\\
G_Q(z,t)&\equiv&\sum_{n=0}^\infty {t^n\over n!}Q_n(z)=(1-2t)^{-1/2}\,
{e^{z\over4(1-2t)}\,k_{-1}\left({z\over4(1-2t)}\right)\over
e^{z/4}\,k_{-1}(z/4)}.
\label{e42}
\end{eqnarray}

The polynomials $P_n(z)$ and $Q_n(z)$ obey identical-looking orthogonality
and normalization conditions
\begin{equation}
\int_{-\infty}^\infty dx\,W_P(x)P_m(x)P_n(x)=
\int_{-\infty}^\infty dx\,W_Q(x)Q_m(x)Q_n(x)=
(2n-1)!!\,(2n+1)!!\,\delta_{m,n},
\label{e43}
\end{equation}
where $(-1)!!=1$. Note that the weight functions $W_P(x)$ and $W_Q(x)$ are {\sl
real} and {\sl positive} and are expressible as principal-part integrals:
\begin{eqnarray}
W_P(x)&=&\int_{-\infty}^x ds\,{\cal P}\int_{-\infty}^\infty {dt\over t-s}
\ln\left[\sqrt{-t\over2\pi}\,e^{-t/4}\,K_1(-t/4)\right],\nonumber\\
W_Q(x)&=&\int_{-\infty}^x ds\,{\cal P}\int_{-\infty}^\infty {dt\over t-s}
\ln\left[\sqrt{-t\over2\pi}\,e^{-t/4}\,K_0(-t/4)\right].
\label{e44}
\end{eqnarray}
The moments of the weight functions $W_P(x)$ and $W_Q(x)$ give rise to the
following lovely continued-fraction expansions:
\begin{eqnarray}
\sum_{n=0}^\infty a_n^{(P)}t^n&=&
\displaystyle{1\over 1-
\displaystyle{3t\over 1-
\displaystyle{t\over 1-
\displaystyle{5t\over 1-
\displaystyle{3t\over 1-
\displaystyle{7t\over 1-
\displaystyle{5t\over 1-\cdots}}}}}}},\nonumber\\
\sum_{n=0}^\infty a_n^{(Q)}t^n&=&
\displaystyle{1\over 1-
\displaystyle{t\over 1-
\displaystyle{3t\over 1-
\displaystyle{3t\over 1-
\displaystyle{5t\over 1-
\displaystyle{5t\over 1-
\displaystyle{7t\over 1-\cdots}}}}}}}.
\label{e46}
\end{eqnarray}

\noindent{\bf Special case $N=1$ (Airy functions).} For this case the
eigenvalues $E$ in (\ref{e19}) are $E=\pm3,~\pm9,~\pm15,~\pm21,~\ldots$ and the
eigenfunctions $y_n(x)$ in (\ref{e20}) are written in terms of Airy functions
combined with polynomials. For negative eigenvalues we have
\begin{equation}
y_{-n-1}(x)=2^{-1/3}\,x\,{\rm Ai}(2^{-2/3}x^2)P_n(4x^3)+{\rm Ai}'(2^{-2/3}x^2)
Q_n(4x^3)\qquad(n\geq0),
\label{e47}
\end{equation}
and for positive eigenvalues we have
\begin{equation}
y_n(x)\equiv y_{-n-1}(-x)\qquad(n\geq0).
\label{e48}
\end{equation}
The polynomials $P_n$ and $Q_n$ are functions of the variable $z=4x^3$. The
first few such polynomials are
\begin{eqnarray}
P_0(z)&=&1,\nonumber\\
P_1(z)&=&z+7,\nonumber\\
P_2(z)&=&z^2+25z+91,\nonumber\\
P_3(z)&=&z^3+55z^2+698z+1729,\nonumber\\
P_4(z)&=&z^4+97z^3+2685z^2+22970z+43225,\nonumber\\
Q_0(z)&=&1,\nonumber\\
Q_1(z)&=&z+5,\nonumber\\
Q_2(z)&=&z^2+23z+55,\nonumber\\
Q_3(z)&=&z^3+53z^2+602z+935,\nonumber\\
Q_4(z)&=&z^4+95z^3+2505z^2+18790z+21505.
\label{e49}
\end{eqnarray}

The polynomials $P_n(x)$ and $Q_n(x)$ satisfy the recursion relations
(\ref{e25}), the coupled second-order differential equations (\ref{e27}),
and the coupled differential relations (\ref{e28}) with $N=1$. The generating
functions $G_P(z,t)$ and $G_Q(z,t)$ and the integral representations of the
weight functions $W_P$ and $W_Q$ are obtained by setting $N=1$ in (\ref{e29a}),
(\ref{e29b}), (\ref{e31}), and (\ref{e32}). The polynomials $P_n(z)$ and
$Q_n(z)$ obey identical-looking orthogonality and normalization conditions
\begin{equation}
\int_{-\infty}^\infty dz\,W_P(z)P_m(z)P_n(z)=
\int_{-\infty}^\infty dz\,W_Q(z)Q_m(z)Q_n(z)=
{3\over\pi}6^{2n}\Gamma\left(n+{5\over6}\right)\Gamma\left(n+{7\over6}\right)
\,\delta_{m,n}
\label{e53}
\end{equation}
and the moments of the weight functions $W_P(x)$ and $W_Q(x)$ have the following
continued-fraction expansions:
\begin{eqnarray}
\sum_{n=0}^\infty a_n^{(P)}t^n&=&
\displaystyle{1\over 1-
\displaystyle{7t\over 1-
\displaystyle{5t\over 1-
\displaystyle{13t\over 1-
\displaystyle{11t\over 1-
\displaystyle{19t\over 1-
\displaystyle{17t\over 1-\cdots}}}}}}},\nonumber\\
\sum_{n=0}^\infty a_n^{(Q)}t^n&=&
\displaystyle{1\over 1-
\displaystyle{5t\over 1-
\displaystyle{7t\over 1-
\displaystyle{11t\over 1-
\displaystyle{13t\over 1-
\displaystyle{17t\over 1-
\displaystyle{19t\over 1-\cdots}}}}}}}.
\label{e55}
\end{eqnarray}

There is an interesting connection between the moments $a_n^{(Q)}$ and the
combinatorial numbers $C_{2n}^{[3]}$, which represent the sum of the
symmetry numbers of the $2n$-vertex connected vacuum graphs in a $\phi^3$
quantum field theory \cite{BM}:
\begin{equation}
a_n^{(Q)}=6n C_{2n}^{[3]} 4^n.
\label{e56}
\end{equation}

\section*{ACKNOWLEDGMENTS}
\label{s6}
CMB wishes to thank A.~Voros and P.~Dorey for interesting discussions and the
CEA, Service de Physique Th\'eorique de Saclay for their hospitality. We also
thank the U.S.~Department of Energy for financial support.

\end{document}